\documentclass{elsart}  
\usepackage{amssymb,amsmath,cite,epsfig, psfrag} 
\newcommand{\be}{\begin{equation}}  
\newcommand{\ee}{\end{equation}}  
\newcommand{\bea}{\begin{eqnarray}}  
\newcommand{\eea}{\end{eqnarray}}  
\newcommand{\bean}{\begin{eqnarray*}}  
\newcommand{\eean}{\end{eqnarray*}}

\newcommand{\gapproxeq}{\lower  
.7ex\hbox{$\;\stackrel{\textstyle >}{\sim}\;$}}  
\newcommand{\lapproxeq}{\lower  
.7ex\hbox{$\;\stackrel{\textstyle <}{\sim}\;$}}

\newcommand{\la}{\langle}  
\newcommand{\ra}{\rangle}  
  
\newcommand{\bc}{\begin{center}}  
\newcommand{\ec}{\end{center}}  
\newcommand{\btab}{\begin{tabular}}  
\newcommand{\etab}{\end{tabular}}

\def\qq{$ q\bar q $}

\def\10bar{$\mathbf{\overline{10}}$}
\def\3bar{$\bar {\hbox{\bf 3}}$}    
\begin{document}  
  
\begin{frontmatter}
\title{The $J=\frac{3}{2}$ $\Theta^*$ partner to the $\Theta(1540)$ baryon}
\author{J.J.Dudek},
\ead{dudek@thphys.ox.ac.uk}
\author{F.E.Close}
\ead{f.close@physics.ox.ac.uk}
\address{ Department of Physics - Theoretical Physics, University of Oxford,\\
1 Keble Rd., Oxford OX1 3NP, UK}
\begin{abstract}  
  If the exotic baryon $\Theta(1540)$ is $udud\bar{s}$ with $J^P =
  \frac{1}{2}^+$, we predict that there is a \10bar 
  with $J^P = \frac{3}{2}^+$ containing a $\Theta^*(1540-1680)$.  
  The width $\Gamma(\Theta^* \to KN)$ is
  at least a factor of three larger than $\Gamma(\Theta)$.  The
  possibilities of $\Theta^* \to KN\pi$ or $\Theta \gamma$ via $M1$
  and $E2$ multipoles are discussed.

\end{abstract}  

\end{frontmatter}

A major plank in establishing the constituent quark model was the
absence of baryons with strangeness +1. The announcement of such a
particle, $\Theta(1540)$, and with a narrow width is therefore
startling\cite{leps}, though there is still some debate as to its
existence\cite{dzierba,fec03}. It is thus important to seek further
evidence of such hadrons in order to isolate the underlying dynamics
of strong QCD. We show here that if $\Theta$ is $udud\bar{s}$ with
$J^P=1/2^+$, then the correlations among QCD forces necessarily imply
there be $\Theta^*$, $J^P=3/2^+$, which is probably only a few tens of
MeV more massive.


When the proton is viewed at high resolution, as in inelastic electron
scattering, its wavefunction is seen to contain configurations where
its three ``valence" quarks are accompanied by further \qq~ in its
``sea". The three quark configuration is thus merely the simplest
required to produce its overall positive charge and zero strangeness.
The question thus arises whether there are baryons for which the
minimal configuration cannot be satisfied by three quarks.  The
$\Theta$ would be an example; the positive strangeness requires an
$\bar{s}$ and $qqqq$ are required for the net baryon number, making
what is known as a ``pentaquark" as the minimal ``valence"
configuration.

Hitherto unambiguous evidence for such states in the data has been
lacking; their absence having been explained by the ease with which
they would fall apart into a conventional baryon and a meson with
widths of many hundreds of MeV.  It is perhaps this feature that
creates the most tantalising challenge from the perspective of QCD:
why does $\Theta$ have width below 10MeV, perhaps no more than
1MeV\cite{1mev}.

There is a considerable literature that recognises that $ud$ in colour
\3bar with net spin 0 feel a strong attraction\cite{scalardiq,dgg}.
Such attraction between quarks in the color \3bar channel halves their
effective charge, reduces the associated field energy and can be a
basis of color superconductivity in dense quark matter\cite{wilc}.  It
has been suggested\cite{kl1,jw} that such correlations might even
cause the $S$-wave combination to cluster as $[udu][d\bar{s}]$ which is
the $S$-wave KN system, while the $P$-wave positive parity exhibits a
metastability such as seen for the $\Theta$ and enables some contact
with the Skyrme model\cite{skyrme} where $\Theta$ is a member of a
\10bar with $J^P = 1/2^+$.  There is the implied assumption
that such a ``diquark" may be compact, an effective boson
``constituent", which is hard to break-up.  We denote this $[(ud)_0]$,
the subscript denoting its spin, and the [ ] denoting the compact
quasiparticle.  Ref.\cite{jw} considers the following subcluster for
the pentaquark: $[(ud)_0][(ud)_0]\bar{s}$.  Ref.\cite{kl1} by contrast
assumes that the $[(ud)_0]$ seed is attracted to a strongly-bound
``triquark" $[(ud)_1\bar{s}]$.

  
More generally, in any pentaquark model with positive parity, angular
momentum is required to annul the intrinsic negative parity of the
$\bar{q}$. Coupling $L=1$ with $\vec{S}_{\bar{q}}$ thus implies that
both $1/2^+$ ($\Theta$) and $3/2^+$ ($\Theta^*$)
exist.  The mass gap $\Delta m (\Theta^* - \Theta)$ is determined by
the strength of the spin-orbit forces within the pentaquark. No
estimate of these exists in the literature to the best of our
knowledge.  This is the issue that we address here. We shall argue
that $\Delta m (\Theta^* - \Theta) \leq m_{\pi}$ such that $\Theta^*
\to \Theta \gamma$ transitions via $M1$ and $E2$ radiation, $\Theta^* \to
KN$ and $\Theta^* \to KN\pi$ are the only allowed decay channels.

Spin orbit forces, Thomas precession and Wigner rotation effects
transforming as $\la \mathbf{L.S} \ra$ are calculated to be
individually sizable in \qq~ and $qqq$ states.  For \qq~ there is a
cancellation among these effects arising from the short-range vector interaction
(single gluon exchange) and long range confinement (assumed to be
scalar)\cite{dalitz,dgg}.  For $qqq$ baryons the situation
is more subtle\cite{dalitz,gromes} and appears to violate
Galilean invariance. Ref\cite{dalitz,gromes} showed this violation to
be illusory and arises because the {\bf p}=0 frame of the two quarks
experiencing the {\bf L.S} interaction is not in general the same as
the overall {\bf P}=0 frame of the $qqq$ three-body system.
Calculations performed in the {\bf P}=0 frame give the correct answer;
in other frames further Wigner rotations would be required, which
transform like {\bf L.S} and contribute to the whole.

For pentaquark systems with tightly clustered {\em scalar} diquarks (as in
refs.\cite{kl1,jw}) the pattern of {\bf L.S} and Thomas precession
will differ radically from the systems \qq~ and $qqq$ where all
pairwise interactions are between fermions. We summarise the results
here and give more details elsewhere.

 \section*{Spin-Orbit Splitting}
 
 We consider the spin-orbit splitting by analogy to that used with
 some success in the conventional meson and baryon sector.
 Conventionally we would consider the non-relativistic reduction of
 the exchange of a particle having some arbitrary propagator between
 two spin-1/2 quarks leading to a Breit-Fermi Hamiltonian. Such a
 Hamiltonian will contain the binding potentials and relativistic
 corrections which include spin-orbit terms. The particular form of
 the spin-orbit terms is fixed by the exchange propagator and the
 Lorentz transformation property of the vertices.  Phenomenological
 success has come from using a combination of a scalar confining
 potential $V_S(r)=b r$ and a vector ($\gamma^\mu$) one gluon exchange
 with $V_V(r) = -\frac{2}{3} \frac{\alpha_S}{r}$ (between two {\bf
   3}'s coupled to a $\mathbf{\overline{3}}$, there is an extra factor
 of $2$ between a {\bf 3} and a $\mathbf{\overline{3}}$ coupled to a
 singlet).
  
 For quarks of mass $m$, the pairwise spin-orbit interaction in a
 vector potential takes the form\cite{dalitz}
  \begin{equation} 
H^V_{SO}(ij) =\frac{1}{4m^2
   r_{ij}}\frac{dV_V}{dr_{ij}}\left(3\vec{L}_{ij}.\vec{S}_{ij}^+ -
   \vec{K}_{ij}.\vec{S}_{ij}^- \right)
  \label{vso}
 \end{equation} 
 where $\vec{L}_{ij} \equiv \vec{r}_{ij} \times (\vec{p}_i -
 \vec{p}_j)$; $\vec{K}_{ij} \equiv \vec{r}_{ij} \times (\vec{p}_i +
 \vec{p}_j)$ and $\vec{S}_{ij}^{\pm} \equiv (\vec{\sigma}_i \pm
 \vec{\sigma}_j)/2$. The analogue for a scalar potential is
\begin{equation} 
H^S_{SO}(ij) = -\frac{1}{4m^2
    r_{ij}}\frac{dV_S}{dr_{ij}}\left(\vec{L}_{ij}.\vec{S}_{ij}^+ +
    \vec{K}_{ij}.\vec{S}_{ij}^- \right)
  \label{sso}
\end{equation} 
For \qq~ with $\mathbf{P}=0$ one has $\mathbf{K} \equiv 0$ and there is a cancellation
between $V_V$ and $V_S$ contributions from the {\bf L.S} terms leading
to a small net spin-orbit splitting, in line with data. For $qqq$ the
{\bf K.S} terms appear to violate Galilean-invariance. This has caused
them to be ignored in some treatments of the spin-orbit splittings of
baryons\cite{isgurkarl}.  The resolution of this involves the great
care necessary when separating c.m. and internal coordinates for
particles with spin\cite{dalitz} the details of which go beyond the
present paper.
  
The essential result of refs.\cite{dalitz} is that the interaction in
eqs.(\ref{vso},\ref{sso}) is correct if applied in the overall
rest-frame $\mathbf{P}=0$ for an $N$-body system.  This generalises to
arbitrary systems of spinors and scalars: construct the corresponding
interaction between spin-0 and spin-1/2 objects and apply the
Hamiltonian in the overall $\mathbf{P}=0$ frame.
  
This is our point of departure for the computation of the spin-orbit
energy shifts of a pentaquark in the models of ref\cite{kl1,jw}. We
assume that the interaction does not resolve any quark substructure
within the diquark or triquark so that they can be considered to be
point-like objects. Labeling the scalar by the subscript $0$ and the
fermion by $f$, the spin-orbit term so obtained has the following
form,
\begin{equation}
  H_{SO} = \frac{\vec{\sigma} \cdot \vec{r} \times \vec{p}}{4 m^2} \left( \frac{1}{r} 
  \frac{d V_V}{dr} - \frac{1}{r}\frac{d V_S}{dr} \right) - \frac{\vec{\sigma} \cdot \vec{r} 
  \times \vec{k}}{2 m m_0} \left( \frac{1}{r} \frac{d V_V}{dr}\right),
\label{hso}
\end{equation}
where $\vec{r} = \vec{r}_{f} - \vec{r}_0$, $\vec{p}=\vec{p}_{f}$,
$\vec{k} = \vec{p}_0$ and $\vec{\sigma}$ is the vector of Pauli spin
matrices which act on the spin-1/2 state-vector.

\section*{Karliner-Lipkin model}

The KL model\cite{kl1} has the $\Theta$ as an effective 2-body system
consisting of a scalar diquark and a spin-1/2 triquark in a relative
$P$-wave. In their paper the masses are quoted as 720 MeV and 1260 MeV
but this does not include the spin-spin interaction energy that binds
them. The $\Theta$ and the $D_s$ composite systems then have roughly
the same reduced mass and are then assumed\cite{kl1} to be bound by
the same (QCD) dynamics. We exploit this analogy and set the
diquark-triquark binding potential to be $V_V(r) = -\frac{4}{3}
\frac{\alpha_S}{r}$, $V_S(r) = br$ with $b=0.18 \mathrm{GeV}^2$ and $\alpha_S \sim 0.5$ (which gives a good fit to the hyperfine shifts of \qq~ and $qqq$)\cite{dgg}. In this potential the $P$-wave $D_s$ mesons
can be described by a variational harmonic oscillator wavefunction,
$R(r) \sim r \exp{- \beta^2 r^2 /2}$ with $\beta \sim 0.4
\mathrm{GeV}$, which reproduces the results of Godfrey and
Isgur\cite{gi} to $\sim$10\%.

With $\vec{r} = \vec{r}_{\mathrm{tri}} - \vec{r}_{\mathrm{di}}$ the
internal momenta are $\vec{p}=\vec{p}_r$, $\vec{k}=- \vec{p}_r$ and
the orbital angular momentum, $\vec{L} = \vec{r} \times \vec{p}_r$.
The spin-orbit splitting term in eq.(\ref{hso}) becomes
\begin{equation}
  H_{SO} = \frac{\vec{S} \cdot \vec{L}}{2 m_{\mathrm{tri}}^2} 
  \left( \frac{4 \alpha_S}{3} \left\la \frac{1}{r^3} \right\ra 
  \left[1+ 2\frac{m_{\mathrm{tri}}}{m_{\mathrm{di}}}\right] - b \left\la 
  \frac{1}{r} \right\ra \right) \nonumber
\end{equation}
and using $\la L=1 | r^{-3(-1)} | L=1 \ra = \frac{4}{3 \sqrt{\pi}}
\beta^{3(1)}$ this gives a splitting of
\begin{equation}
  \Delta E(\Theta^* - \Theta) = \frac{1}{\sqrt{\pi} m_{\mathrm{tri}}^2} \left( \frac{4 \alpha_S}{3} \beta^3 \left[1+ 2\frac{m_{\mathrm{tri}}}{m_{\mathrm{di}}}\right] - b \beta \right).\label{KLsplit}
\end{equation}
Using $m_{\mathrm{di}}=720 \mathrm{MeV}$ and $m_{\mathrm{tri}}=1260
\mathrm{MeV}$\cite{kl1}
\begin{equation}
 \Delta E(3/2 - 1/2) = (63 \mathrm{MeV})_V + (-25 \mathrm{MeV})_S = 38 \mathrm{MeV}, \nonumber
\end{equation}
where a cancellation between vector and scalar terms is observed much
as in the conventional \qq~ and $qqq$ states. Each term is
individually small due to the large $m_{\mathrm{di}}$ and
$m_{\mathrm{tri}}$.

We investigate how much this splitting could be increased if we
minimise these ``constituent" masses by (i) including the subtracted
hyperfine energy internal to the diquark and triquark (ii) maximising
the orbital excitation energy to 500MeV.  If we insist that in general
the sum of the diquark and triquark masses is 1540 MeV less a $P$-wave
energy under 500 MeV, we cannot get a spin-orbit splitting greater
than about 150 MeV which is still below threshold for the strong
channels $\Theta^* \to \Theta \pi\pi$ or $K\Delta$ (which is forbidden
for $\Theta^*\supset$\10bar).

\section*{Jaffe-Wilczek model}

\begin{figure} 
\begin{center} 
  \psfragscanon \psfrag{R}[]{$\frac{1}{\sqrt{2}}\vec{\rho}$}
  \psfrag{L}[]{$\sqrt{\frac{3}{2}}\vec{\lambda}$} \psfrag{1}[]{\bf 1}
  \psfrag{2}[]{\bf 2} \psfrag{3}[l]{\bf 3}
  \includegraphics[width=3in]{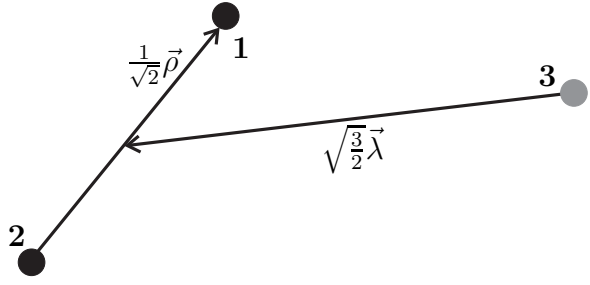}
\caption{3-body variables \label{3body}}  
\end{center}
\end{figure}

In the JW model\cite{jw} the $\Theta$ is effectively a 3-body system
of two identical scalar diquarks in a $P$-wave and an antiquark in a
relative $S$-wave. The color structure is anti-baryon-like ${\bf
  \bar{3}} \otimes {\bf \bar{3}} \otimes {\bf \bar{3}} = {\bf 1}$. The
Breit-Fermi Hamiltonian for this system is obtained by summing over
exchanges between the three bodies pairwise.  Label the diquarks
$1,2$ (mass $m_0$); the antiquark $3$ (mass $m$); and define(see
Fig.\ref{3body})
\begin{eqnarray}
  \vec{r}_{1,2} &=& \vec{R} + \sqrt{\frac{3}{2}} \frac{m}{2m_0 + m} \vec{\lambda} \pm 
  \frac{1}{\sqrt{2}} \vec{\rho}\nonumber \\
  \vec{r}_3 &=& \vec{R} - \sqrt{\frac{3}{2}} \frac{2m_0}{2m_0 + m} \vec{\lambda}. \nonumber
\end{eqnarray} 
In the $\Theta$ rest frame the internal momenta are
\begin{eqnarray}
  \vec{p}_{1,2} &=& \frac{1}{\sqrt{6}} \vec{p}_\lambda \pm \frac{1}{\sqrt{2}} \vec{p}_\rho \nonumber \\
  \vec{p}_3 &=& - \sqrt{\frac{2}{3}} \vec{p}_\lambda.\nonumber
\end{eqnarray}
Hence the one unit of orbital angular momentum between the diquarks is
with respect to the $\vec{\rho}$ variable, $L_\rho =1$ whereas the
$\vec{\lambda}$ variable is in an $S$-wave, $L_\lambda =0$. (This is
opposite to the $L=1$ $qqq$ where the symmetry exposes the excitation
of the $\lambda$ oscillator). The spin-orbit term is then (note,
exchanges between the scalar diquarks do not contribute here)
\begin{eqnarray}
  H_{SO} &=& \left\{ \left( 1 + \frac{m}{m_0} \right) \Sigma_V - \Sigma_S \right\} \frac{\vec
  {\sigma} \cdot \vec{\lambda} \times \vec{p}_\lambda}{4 m^2} 
+ \Sigma_V \frac{\vec{\sigma} \cdot \vec{\rho} \times \vec{p}_\rho}{4 m m_0}  \nonumber \\
&+&\frac{1}{\sqrt{3}} \left\{ \left(1 + \frac{m}{m_0} \right) \Delta_V - \Delta_S \right\}  
\frac{\vec{\sigma} \cdot \vec{\rho} \times \vec{p}_\lambda}{4 m^2} 
+\sqrt{3} \Delta_V \frac{\vec{\sigma} \cdot \vec{\lambda} \times \vec{p}_\rho}{4 m m_0} ,
\end{eqnarray} 
where $\Delta_{V,S} = \frac{1}{r_+} \frac{d V_{V,S}}{d r_+} -
\frac{1}{r_-} \frac{d V_{V,S}} {d r_-}$, $\Sigma_{V,S} = \frac{1}{r_+}
\frac{d V_{V,S}}{d r_+} + \frac{1}{r_-} \frac{d V_{V,S}} {d r_-}$,
$\vec{r}_{\pm} = \vec{r}_3 - \vec{r}_{1,2}$ and $r_{\pm} =
|\vec{r}_{\pm}| = \frac{1} {\sqrt{2}} \sqrt{3 \lambda^2 \pm 2 \sqrt{3}
  \vec{\lambda}\cdot\vec{\rho} + \rho^2}$. We again use vector
one-gluon-exchange and now a general scalar potential to describe the
binding.

The scalar interaction does not contribute to the spin-orbit splitting
in this model.  This is because terms featuring $\vec{p}_\lambda$ are
trivially zero - $\vec{p}_\lambda$ acting on the $L_\lambda=0$
wavefunction is proportional to $\vec{\lambda}$, so that the first
term is $\vec{\lambda} \times \vec{\lambda}$. The $\vec{\rho} \times
\vec{p}_\lambda$ term $\sim \vec{\rho} \times \vec{\lambda}$,
integrating over the direction of $\vec{\lambda}$ gives something in
the direction of $\vec{\rho}$ and hence $\vec{\rho} \times \vec{\rho}
= 0$.

\begin{figure} 
\begin{center} 
  \psfragscanon \psfrag{R}[]{$\frac{1}{\sqrt{2}}\vec{\rho}$}
  \psfrag{L}[]{$\sqrt{\frac{3}{2}}\vec{\lambda}$} \psfrag{1}[b]{\bf 1}
  \psfrag{3}[]{\bf 3} \psfrag{Rlessl}[l]{$\rho < \sqrt{3} \lambda$}
  \psfrag{Rgreaterl}[l]{$\rho > \sqrt{3} \lambda$}
  \includegraphics[width=5in]{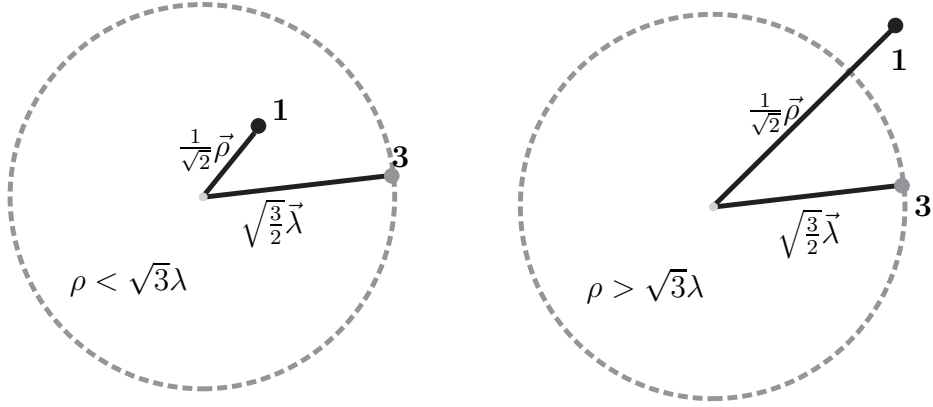}
\caption{after averaging over directions of $\vec{\lambda}$, the diquark feels no force when inside the sphere and feels a force in the direction of $\vec{\rho}$ when outside. \label{gauss}}  
\end{center}
\end{figure}

The two terms proportional to $\vec{p}_\rho$, which are driven by the
vector interaction, are non-zero only when $\sqrt{3} \lambda \leq
\rho$ which is a consequence of Gauss's law. They are proportional to
the force on the diquarks due to the antiquark which is only non-zero
when the diquark is outside the spherical shell obtained by averaging
over directions of $\vec{\lambda}$ with fixed $\lambda$ - see
Fig.\ref{gauss}. This is rather similar to the baryon model considered
in \cite{isgurls}.  Thus although there is no cancellation between
vector and scalar in this model, the spatial restriction to the
spherical shell defined by $\rho$ enfeebles the total {\bf L.S}
contribution here.

Specifically, after integration over the direction of $\vec{\lambda}$
these terms combine to give a spin-orbit term for $\rho > \sqrt{3}
\lambda$,
\begin{equation}
\frac{2 \sqrt{2}}{m m_0} \frac{2 \alpha_S}{3} \frac{1}{\rho^3} \vec{S} \cdot \vec{L}_\rho, \nonumber
\end{equation} 
and a splitting
\begin{equation}
  \Delta E(3/2 - 1/2) = \frac{2 \sqrt{2} \alpha_S}{m m_0} \left \la \frac{1}{\rho^3}
  \theta(\rho-\sqrt{3} \lambda) \right \ra \nonumber
\end{equation} 
where only the vector interaction contributes.

Jaffe and Wilczek do not explicitly state the masses of their
diquarks. The standard {\bf S.S} interaction would give $m_0 \sim
500$MeV, but this is hard to confront with a $\Theta(1540)$ containing
two diquarks, together with $m_s$ and a $P$-wave excitation energy
($\omega$) as well.  Further, one should ensure that
$[ud]_0[ud]_0[ud]_0$ with two $P$-waves, is not more stable than the
deuteron. This requires that $\omega \gapproxeq 450$MeV. Such a result
is at least consistent with the (anti)baryon spectrum, which has the
same internal colour arrangement and by assumption similar binding
dynamics for which the energy gap between $(m_N + m_{\Delta})/2$ and
the negative parity $N^*(1520-1750)$ is $\sim 500$MeV.  This leaves
$\sim 1-1.1$GeV to be shared between two diquarks and the $\bar{s}$.
With $m_s \sim 450 \mathrm{MeV}$ we thus assign a mass of $\sim 350
\mathrm{MeV}$ to each diquark. This is the minimum we can tolerate,
and even with this we shall find that the $\Theta^* -\Theta$ mass gap
is only tens of MeV; any larger mass for the diquarks would reduce it
even more.

We can make an estimate of the $\Theta^{(*)}$ spatial wavefunction
from baryon measurements.  The $\Lambda-\Sigma$ splitting can be used
to find the harmonic oscillator parameter, $\alpha_\rho \approx 400
\mathrm{MeV}$;
\begin{equation}
M(\Sigma^0)-M(\Lambda^0) \approx 77 \mathrm{MeV} = \frac{16 \pi 
\alpha_S}{9 m}\left(\frac{1}{m}-\frac{1}{m_s}\right) \frac{\alpha_\rho^3}{\pi^{3/2}}. \nonumber
\end{equation}
This gives a level spacing $\omega (L=0,L=1) = \alpha_\rho^2/m \sim
485 \mathrm{MeV}$ in good agreement with the data\cite{pdg}.

The radial integrals can be performed with the approximate harmonic
oscillator wavefunctions giving
\begin{equation}
\Delta E(3/2 - 1/2) = \frac{2 \sqrt{2} \alpha_S}{m m_0} \left(\frac{4}{3 \sqrt{\pi}} \frac{k^3}
{(3+k^2)^{3/2}}\right) \alpha_\rho^3 \label{JWsplit}
\end{equation}
where $k=\left(\frac{3m}{2m_0 +m}\right)^{1/4}$. With $\alpha_\rho
\sim 400 \mathrm{MeV}$ this gives $\Delta E \sim 35, 65 \mathrm{MeV}$
with $m_0 = 500,300 \mathrm{MeV}$ respectively.

In all cases considered here, the spin-orbit excitation energy is
plausibly small on the scale of the mass gap between scalar and vector
diquark, which is expected to be $O(200)$MeV.  That this is so is an
explicit assumption here, and implicitly assumed in ref.\cite{jw} for
the required correlations to obtain. Thus we conclude that the first
$\Theta^*$ will be the {\bf L.S} state discussed here; states where
the clusters are internally excited, if any, will be at higher masses.
However, if the clusters can be resolved, there is the possibility of
dilution of the present effects, or of introducing non-zero
contributions from tensor forces for example. It cannot be excluded
that these could conspire to reduce the net mass gap, even that the
$\Theta^*$ and $\Theta$ are degenerate within the present resolution
of the experimental data.

\section*{Other states}

Ref\cite{jw} consider $[(ud)_0][(ud)_0]\bar{Q}$ with $Q \equiv u,d$.
These are expected to lie $O(100-150)$MeV below the $\Theta$, and may
be identified with the $P_{11}(1440)$ ``Roper" resonances\cite{pdg}.
The {\bf L.S} forces then imply a $3/2^+$ partner at $\sim
1.5-1.6$GeV, for which there is no evidence\cite{pdg}. However, this
\10bar -{\bf 8} mixture can couple to conventional $qqq$ (nucleon) by
\qq~ annihilation, which may explain both its large width and possibly
lead to large mass shifts, and unobservably large widths of any
$3/2^+$ partner.  For the $[(ud)_0][(us)_0]\bar{s}$ state, the
mass gap to the $3/2^+$ partner is predicted to be $\sim
O(10)$MeV. This tantalisingly is in accord with the
$P_{11}(1710)-P_{13}(1720)$ pair.
 
The mixed messages here may indicate that the concept of constituent
pentaquarks is meaningful only for manifestly exotic combinations. The
dynamics of {\bf L.S} may therefore be probed also by the exotic
$\Xi^+$, $\Xi^{--}$ and the heavy flavour analogues $[ud][ud]\bar{Q}$,
with $Q \equiv c,b$.  We predict $\Delta m(\Xi^{*} - \Xi) \sim
30-50$MeV.  The reason that this is similar to $\Delta
m(\Theta^*-\Theta)$ is due to the effect of the heavier $s$ mass being
``diluted" within clusters, and the $\bar{s}$ being replaced by the
lighter $\bar{d}(\bar{u})$.  The splitting scales as the
inverse constituent mass for large $m_Q$(eqns.\ref{KLsplit}, \ref{JWsplit}) and hence the splittings for
$\Theta_c^* - \Theta_c$ are at most a few MeV.

 \section*{Decays of $\Theta^*$}
 
 The ``natural" width of a $P$-wave $\Theta$ resonance 100MeV above $NK$
 threshold is of order 200MeV\cite{jw,maltman}. However, this has not
 yet taken into account any price for recoupling colour and
 flavour-spin to overlap the $(ud)(ud)\bar{s}$ onto $NK$ colour
 singlets such as $uud$ and $d\bar{s}$.  In amplitude, starting with
 the Jaffe-Wilczek configuration, the colour recoupling costs
 $\frac{1}{\sqrt{3}}$ and the flavour-spin to any particular channel
 (e.g. $K^+n$) costs a further $\frac{1}{2\sqrt{2}}$.
 If in addition we suppose that spatially the constituents then fall
 apart, only the $L_z=0$ piece of the wavefunction contributes, which
 implies a further suppression in amplitude of $\sqrt\frac{1}{3}$ for
 the $L=1\otimes S=\frac{1}{2} \to J=\frac{1}{2}$. Thus a total
 suppression in rate of up to two orders of magnitude may be
 accommodated in such pictures.
 
 For the $\Theta^*$ similar arguments obtain, though $\Gamma_\mathrm{TOT}$ is
 now expected to be larger as (i) the spatial Clebsch is now
 $\sqrt\frac{2}{3}$ instead of $\sqrt\frac{1}{3}$; (ii) the phase
 space $\sim q^3$ gives a factor $1.8-3.5$. So $\Gamma(\Theta^*(1600))
 \sim 3.6 \Gamma(\Theta)$, rising to $\sim 7 \Gamma(\Theta)$ if $m
 \sim 1700$MeV.  However, if the decay involves tunneling, which is
 exponentially sensitive to the difference between the barrier height
 and the kinetic energy of the state, this could significantly enhance
 the width of the $\Theta^*$\cite{maltman}.

\begin{figure} 
\begin{center} 
  \psfragscanon \psfrag{N}[]{$N$} \psfrag{Th}[l]{$\Theta$}
  \psfrag{Thstar}[l]{$\Theta^*$} \psfrag{G}[]{$\gamma$}
  \psfrag{K}[l]{$K$} \includegraphics[width=2in]{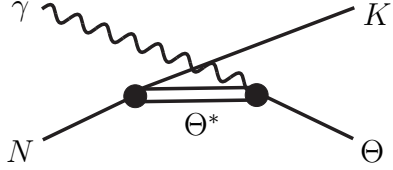}
\caption{u-channel production diagram. \label{uchannel}}  
\end{center}
\end{figure}

 With the
 mass gaps as predicted here, $\Theta^* \to \Theta \pi\pi$ is kinematically forbidden; the possibility of $\Theta^* \to KN\pi$
 emerges, but the phase space for a non-resonant three body decay may
 prevent this being a large branching ratio (unless the $N\pi$ is
 enhanced by the tail of a pentaquark $P_{11}(1440)$). The
 $b.r(\Theta^* \to \Theta \gamma)$ is $\leq 10^{-3}$ due to the
 restricted phase space. However the ratio of $M1$ amplitudes for $\gamma \Theta \to \Theta^*$ and $\gamma N \to \Delta$, with momentum factors removed, is $\frac{m_u}{3}\left(\frac{1}{m_s} + \frac{1}{m_0}\right) \sim 0.4$ and so it
 is possible that this feeds a significant $\gamma N \to \Theta K$ via
 $\Theta^*$ exchange in the $u$-channel (Fig \ref{uchannel}).  Whereas the $\Delta \to N
 \gamma$ in the constituent quark model has $E2 \equiv 0$ due to the
 $L=0$ internal structure of these baryons, the $\Theta^* \to \Theta
 \gamma$ has $E2$ arising from the internal orbital degree of freedom.

 In summary pentaquark models of $J^P = 1/2^+$ $\Theta(1540)$ imply
 there be a copy of the \10bar (and {\bf 8}) containing
 $\Theta^*$ with $J^P=3/2^+$ and within tens of MeV of it. We
 advocate searching for this in $KN$ or $KN\pi$ final states. There is
 also the possibility that $\Theta^*$ is actually the 1540 state
 already observed, and that the true $1/2^+$ state lies around
 1500MeV, or that the two states are degenerate within the present
 resolution of the data.  Either of these could explain the conundrum
 of why there is no clear sign of a more prominent narrow structure in
 the $KN$ spectrum above 1540MeV. The possibility that the $\Theta^*$ is broad\cite{maltman} might also explain the present
 data.  An alternative possibility is that the observed $\Theta$ is
 the decay product of some directly produced particle. In the
 photoproduction experiments this could be via the $\Theta^*$ in the
 u-channel. The spectrum of the Skyrme model as described in e.g.\cite{kopel}
 has no place for such a $J^P = 3/2^+$ \10bar multiplet. As such our prediction could
 be an interesting test of models\cite{gloz}.

\bc {\bf Acknowledgments} \ec
  
This work is supported, in part, by grants from the Particle Physics
and Astronomy Research Council, and the EU-TMR program ``Euridice'',
HPRN-CT-2002-00311.

\end{document}